\documentstyle[11pt,aaspp4]{article} 
\newcommand{\etal}{{\it et al.\/}}

\lefthead{Cohen {\it et al.} }
\righthead{Redshift Survey}

\slugcomment{submitted to The Astrophysical Journal}

\begin{document}

\title{Caltech Faint Galaxy Redshift Survey VII:  
Data Analysis Techniques and Redshifts in the Field J0053+1234\altaffilmark{1}}

\author{Judith G. Cohen\altaffilmark{2},
        David W. Hogg\altaffilmark{3,4,5}
        Michael A. Pahre\altaffilmark{2,5,6}
        Roger Blandford\altaffilmark{3},
        Patrick L. Shopbell\altaffilmark{2} \&
        Kevin Richberg\altaffilmark{2}}

\altaffiltext{1}{Based in large part on observations obtained at the
	W.M. Keck Observatory, which is operated jointly by the California 
	Institute of Technology and the University of California}
\altaffiltext{2}{Palomar Observatory, Mail Stop 105-24,
	California Institute of Technology, Pasadena, CA \, 91125}
\altaffiltext{3}{Theoretical Astrophysics, California Institute of Technology,
	Mail Stop 130-33, Pasadena, CA \, 91125}
\altaffiltext{4}{Current Address:  Institute for Advanced Study, Olden Lane, Princeton, NJ \, 08540}
\altaffiltext{5}{Hubble Fellow}
\altaffiltext{6}{Current Address:  Harvard-Smithsonian Center for Astrophysics, 
	60 Garden St., Mail Stop 20, Cambridge, MA \, 02138}

\begin{abstract}
We present the techniques used to determine redshifts and to characterize
the spectra of objects in the Caltech Faint Galaxy Redshift Survey
in terms of spectral classes and redshift quality 
classes.  These are then applied to spectra from an investigation of a 
complete sample of
objects with $K_s<20$~mag in a 2 by 7.3~arcmin$^2$ field at J005325+1234.  
Redshifts were successfully obtained for 163 of the 195 objects in the sample; these redshifts
lie in the range [0.173, 1.44] 
and have a median of 0.58 (excluding 24 Galactic stars).
The sample includes two broad lined AGNs and one QSO.
\end{abstract}


\section{Introduction \label{introduction} }

The Caltech Faint Galaxy Redshift Survey, (henceforth CFGRS),
is designed to measure the properties of field galaxies in the redshift
interval $0.3\lesssim z\lesssim1.3$.  
It uses complete samples to a fixed limiting magnitude in a particular 
bandpass within a small solid angle on the sky.
Spectra are obtained for every object in the sample with the Low
Resolution Imaging Spectrograph (henceforth LRIS) (Oke \etal\ 1995) on
the 10~m Keck Telescope.  
The defining features of this program that distinguish it from existing 
and ongoing or planned surveys, such as 
the CfA1 Survey (\cite{huchra83}),
the Las Campanas Redshift Survey (\cite{shectman96}),
the Canada-France Redshift Survey (henceforth CFRS; Lilly \etal\ 1995a,b\markcite{lilly95a,lilly95b}),
the Autofib survey (\cite{ellis96}),
the Century Survey (\cite{geller97}), 
the ESO slice survey (\cite{vettolani97}),
the Sloan Digital Sky Survey (\cite{gunn93}),
the Deep Survey planned by the Lick group (\cite{koo98})
and the 2dF Survey (\cite{colless98}),
are that it is possible to reach to fainter magnitudes
($K \sim 20$~mag, $R \sim 24$~mag) and that completeness
is emphasized rather than sparse sampling over a large field.  
The closest counterpart to this approach is the work of 
Cowie \etal\ (1996\markcite{cowie96}), although we note that the present
survey covers a larger solid angle at the survey limit.
See Koo \& Kron (1992\markcite{koo92}) and Ellis (1997\markcite{ellis97}) 
for reviews of the subject.

There are four limitations to ``pencil beam'' redshift surveys of the type 
described in the present paper when addressing issues of galaxy evolution.  
The first is that the faintest fluxes
at which sources can be observed in a complete redshift survey are
still over four magnitudes brighter than the fluxes of the faintest
detectable sources in imaging data.  
The second is that the total number of sources and total sky area are both 
limited by available telescope time because the faintest possible 
sources are being studied.  This is particularly
relevant to attempts to characterize large scale structure.
The third is that it is difficult to measure spectroscopic redshifts
in the range $1.3<z<2.3$---the lower limit is set by the
[O~II] line at 3727\AA\ moving out of the optical regime, the 
upper limit by when Ly$\alpha$ enters it.  This 
may be a crucial period in the formation and evolution of normal galaxies (\cite{madau98}).
The final limitation is the difficulty of generalizing the observed properties
in a small field (which may contain unique features) to the properties of the
field galaxy population as a whole. 

This paper presents spectroscopic results from one field located at J005325+1234, 
the central region of which is part of the Medium Deep Survey (\cite{griffiths94})
and among the deepest fields imaged with HST prior to the Hubble Deep Field.
The field measures $2 \times 7.3$~arcmin$^2$ with a statistical 
sample containing 195 
infrared-selected objects complete to $K = 20$~mag, of 
which 24 are spectroscopically confirmed Galactic stars 
and 32 cannot be assigned spectroscopic redshifts. (21 of these have spectra). 
There are 13 additional objects with redshifts (including two more stars)
that are just outside the field 
boundary or are within the field but too faint.

This paper is structured as follows.
The sample is defined in \S\ref{sampledef}, which is followed by a description
of the spectroscopic 
observations and redshift determinations in \S\ref{redshiftdet}.  
The galaxy spectra are assigned to spectroscopic and quality classes, a procedure discussed in \S\ref{specclass}.

\section{Sample Definition \label{sampledef} }

The sample, defined in Pahre \etal\ (1998\markcite{pahre98apjs}), 
is selected by the criteria that $K_s < 20$~mag 
within a $2 \times 7.3$~arcmin$^2$
field centered at 00 53 23.20 +12 33 58 (J2000).  
In an effort to cut down the thermal background of the standard Johnson $K$ filter, 
which has a central wavelength 2.2$\mu$m and FWHM 0.4$\mu$m, there are at least two 
non-standard filters in this same atmospheric window in common use:  the 
$K^{'}$ filter of Wainscoat \& Cowie (1992\markcite{wainscoat92}) and the 
$K_s$ ($K$--short) filter of Persson (private communication) and Skrutskie \etal\ (1999).
The latter is used here, and covers at half maximum transmission the wavelength
range $2.00 < \lambda < 2.32 \mu$m.
Throughout this series of papers, unless otherwise specified, we use $K$ 
(denoting the standard Johnson filter) and $K_s$ (denoting the $K$--short filter) 
interchangeably, although the appropriate $K_s$ transmission curve is used when
deriving the $k$--corrections in Cohen et al (1998).

This sample contains 195 objects, and we refer to this as the ``main'' sample.
It includes 24 spectroscopically confirmed Galactic stars.  
The remaining 171 objects 
constitute the ``galaxy'' sample. We were able to assign redshifts to 139
of these galaxies and this is the ``redshift'' sample.
Redshifts were also obtained for several objects that either had 
$K > 20$~mag or lie just outside the boundary of the sample.
A few objects which are very faint at both $R$ and $K$ turned up 
serendipitously in slitlets intended for nearby brighter objects.  
Adding these 13 additional objects to the main sample defines the ``total'' 
sample of 208.

The photometric data of Pahre \etal\ (1998\markcite{pahre98apjs}) are 
already corrected for reddening by the ISM in the Galaxy.  
The variation in $E(B-V)$ across the small field of our survey is insignificant.  
As is conventional, we ignore reddening internal to the galaxies themselves 
and from the IGM.

\section{Redshift Determinations \label{redshiftdet} }

\subsection{Data Acquisition and Reduction \label{dataacquire} }

The objects in this sample were observed using the LRIS at the Keck
Observatory beginning in October 1994 and ending in January 1998.
The LRIS detector is a $2048 \times 2048$ pixel Tektronix CCD.
Its dual amplifier readout mode was normally
used to save time.  Observations were carried out even when the
weather was not photometric.  Fifteen slit masks were used.  
Each mask has 25 to 33 slitlets and many objects were observed more
than once.  A few of the brightest galaxies were selected as aids to
align the masks, and hence were observed many times.

Except for the last three masks (the
1997 and 1998 data), the 300 g/mm grating was used with 1.1~arcsec wide
slits.  This produces a scale of 2.46 \AA/pixel and a resolution of 10\AA. 
(This is $\sim$4 times higher spectral resolution than that of the
CFRS.) 
The spectral coverage is $\sim$5000\AA, with the central wavelength
depending on the location of the object on the sky with respect to the
centerline of the slitmask as well as on the grating tilt.  Two 3000
sec exposures were obtained for each of these masks.    

The last three masks, containing most of the faintest
objects in the sample, were observed with the 155 g/mm grating 
with 1.0~arcsec wide slits giving a spectral resolution of
20\AA.  Two or three 1800 sec exposures were
obtained for each of these slitmasks.  Each of these lower dispersion
spectra was slightly shifted in central wavelength (i.e., dithered
spectrally between exposures).  This forces a more complex reduction,
but better signal-to-noise spectra are achieved.  Full spectral
coverage from the UV through 10,000\AA\ is achieved with this grating
for essentially all objects at the price of lower dispersion.

These multi-slit spectra were reduced in a standard way.  The bias
(actually right and left amplifier bias levels) was subtracted.  Then,
the relevant pieces were cut out of each of the multi-slit spectra for
a particular object (and out of the flat field calibration exposure),
cosmic rays were removed, the subsets were flattened and the bowing of
the spectra (the S-distortion) was removed.  When necessary, the
distortion in the vertical direction along the slit which manifests
itself as tilted night sky lines was also removed.

The night sky lines within the spectra themselves were used to define
the wavelength scale; a third order polynomial fit is sufficient.
This means in effect that there are no ``arc'' lines bluer than
5199\AA.  Sometimes that night sky emission line could not be
detected, in which case 5577\AA\ became the bluest night sky line.  Sky
subtraction was accomplished by fitting a second order polynomial to a
range of pixels above and below the object spectrum.  Linear fits were
used when necessary, particularly when the object was near the top
or bottom edge of the slitlet.

The spectra were reduced by JGC using Figaro (\cite{shortridge88}) scripts
and about 20\% were also done by DWH using IRAF\footnote{IRAF is
distributed by the National Optical Astronomy Observatories, which is
operated by the Association of Universities for Research in Astronomy,
Inc.\ (AURA) under cooperative agreement with the National Science
Foundation.}  scripts.  JGC measured all the redshifts
uniformly, including remeasurement and in some cases rereduction of the
spectra done by DWH as well.  The redshifts were measured by visually
determining the wavelengths of spectral features from plots of the
spectra over various wavelength ranges on an interactive display.  The
assignment of rest wavelengths to the observed spectral features was
straightforward for the brighter objects and for those objects with
strong emission lines, but became more difficult for the faint objects
showing only absorption line spectra.  

While a manual redshift
determination may
seem outmoded in light of more automatic techniques such as that 
developed by Glazebrook, Offer, \& Deeley (1998\markcite{glazebrook98}) 
for the 2dF or that of Kurtz \& Mink (1998\markcite{kurtz98}) for the CfA surveys,
in this situation it is more appropriate.  Here the objects
are very faint relative to the background sky, the SNR is low,
the sky subtraction is difficult, the exposures are long (thus increasing
the number of cosmic ray hits), and most importantly we do not fully
understand what features to expect in the spectra of such faint and distant
galaxies.  Construction of a set of templates for a machine based
redshift search from (or even for) this data set
would be impossible.

The spectra have been fluxed using long slit observations of standard
stars from Oke (1990\markcite{oke90}) taken with the same LRIS configuration as was
used for the multi-slits.  The absolute scale of a fluxed spectrum
should be regarded with caution, as the nights were not all
photometric, there are substantial slit losses for the more extended
objects, and slitmask alignment causes an additional light loss which
is variable from mask to mask and from object to object on a given
mask.  However, the {\it shape} of the continuum for a particular object
should be more or less correct, and is often quite useful.

\subsection{Redshift Measurements \label{redshiftdeterminations} }

Redshifts were measured for each object in each of the slitmasks.
Records were also kept of the results for any serendipitous
objects that fell onto slits for other targeted galaxies.  At the end,
the results were compiled, and the very small number
of objects with multiple observations that
showed discrepancies were studied more carefully to resolve them.
This involved summing all the spectra together, examining the
wavelength range of each of the spectra, etc.

The most difficult cases involved objects that show 
no emission features, but only absorption lines
which do not correspond to the standard pattern of Balmer lines,
H+K \ion{Ca}{2} doublet, etc. seen in the region of the 4000\AA\ break.
Since the 4000\AA\ region is redshifted into the strong
night sky emission lines redward of the optical
bandpass for $z > 1$, it was suspected that for such objects $z > 1$.
To deal with
these particular cases, the spectra were compared directly with spectra
of high redshift
galaxies (\cite{cowie95}; \cite{steidel96}),
the sum of spectra of absorption line galaxies in a cluster at $z$ = 1.50
kindly supplied by J.~B.~Oke, as well as the sums of spectra of galaxies
in the higher redshift peaks in the present sample.  The 
library of mean ultraviolet spectral energy distributions for
stellar groups from IUE spectra (\cite{fanelli92})
and those of model galaxies from the grid of Bruzual \&
Charlot (1993\markcite{bruzual93}) were also examined.

The absorption features in the spectra of many of those objects
suspected to be at $z > 1$ are real and repeatable.  The uncertainty
in the redshifts is because of limited knowledge of this spectral
region (for which the 2800\AA\ region is shifted into the optical) and lack
of confidence when there are only two certain features in a spectrum,
both absorption lines, as to what the correct identification should be.
A good example is D0K42 (which has $R_{obs} = 23.12$, 
$R-K = 5.14$), which could have either $z = 0.54$ or the final
adopted value of $z = 1.14$ depending on whether the identification
assigned to the strongest absorption features is H+K of \ion{Ca}{2} or
2800+2850\AA\ (\ion{Mg}{2}+\ion{Mg}{1}).  In this case, the spectral energy distribution of
the fluxed spectra and the absence of a  4000\AA\ jump in the 
$z = 0.54$ case were considered to be persuasive evidence for the
higher value.

The composite quasar spectrum of Francis \etal\ (1991\markcite{francis91}) provided
a template for line identification in the broad emission line objects.

\subsection{Quality Classes \label{qualityclasses} }

We assign a quality class to the redshift of each object to give
an indication of the associated uncertainty.  Table~1 lists the
quality classes (0 through 9) with a brief description of each.
Objects with multiple
features are assigned quality class 1 -- 3 redshifts, depending upon the 
security of the redshift.  Centroiding uncertainties for high
SNR spectra are generally considered to be less than $0.1 \times $~FWHM , but
with the lower SNR prevalent here, we adopt a FWHM as a more
appropriate uncertainty.  (The uncertainty from the wavelength fit itself using
the night sky lines is small by comparison.)  For a feature at
6000\AA\ with $z < 1.5$, a 2 pixel error with the 300 g/mm grating
corresponds to a redshift uncertainty $\le 0.002$.

\placetable{tab1} 

Single emission line spectra are placed in classes 4,5.  The
emission line is always assumed to be 3727\AA\ [\ion{O}{2}].  The CFRS
has adopted a similar strategy (\cite{lilly95b}), although their
situation is more complex as they are more likely to have confusion
between 3727\AA\ emission and H$\alpha$ emission due to the
preponderance of lower redshift galaxies in the CFRS.
In our case, with our relatively broad spectral coverage, the absence
of H$\beta$ or 4959\AA\ emission serves to rule out the possibility
that a single emission line is 5007\AA\ of [\ion{O}{3}].  It cannot be
H$\alpha$ for lines that appear at wavelengths bluer than the rest
wavelength of H$\alpha$, while for redder lines, one would then expect
to see H$\beta$ or the 5007\AA\ [\ion{O}{3}] line in emission.
The only other serious possibility consistent with the absence of features in
optical spectra with broad spectral coverage is Ly$\alpha$;  this 
is a serious possibility for some of the faintest objects in the 
sample, particularly the serendipitous
ones, where only a single emission line from a very faint object falls
by chance onto our slit, cf. Dey \etal\ (1998\markcite{dey98}).
In such cases, broad band colors might be used to look for breaks or
UV dropouts, although we have chosen to assign redshifts on the basis 
of spectroscopy alone. The presence of a few high redshift objects
in our sample should not affect our conclusions which are restricted to the
properties of galaxies with $z<1.5$.

%
%
Objects showing a single break, assumed to be the 4000\AA\ break, are
assigned quality class 8.  Again this is a conservative choice,
since at a modest redshift well within the
range covered by our survey, the break just to the red of the
\ion{Mg}{2} doublet at 2800\AA\ will lie in the optical.  There are
only two objects in this class in the total sample. Quality class 9
is for objects showing a single strong absorption feature which,
because of the shape of the continuum, we interpret as 
\ion{Mg}{2}+\ion{Mg}{1} at
2800+2850\AA\ rather than as Ca~H+K.  There are only three such objects.

Quality class 0 is for objects with very low signal and no redshift.

For future reference, the term ``high quality redshifts'' refers
to those with a quality class indicating a secure $z$, specifically
quality classes 1, 2, 4, or 6.

\begin{figure}
\epsscale{0.7}
\plotone{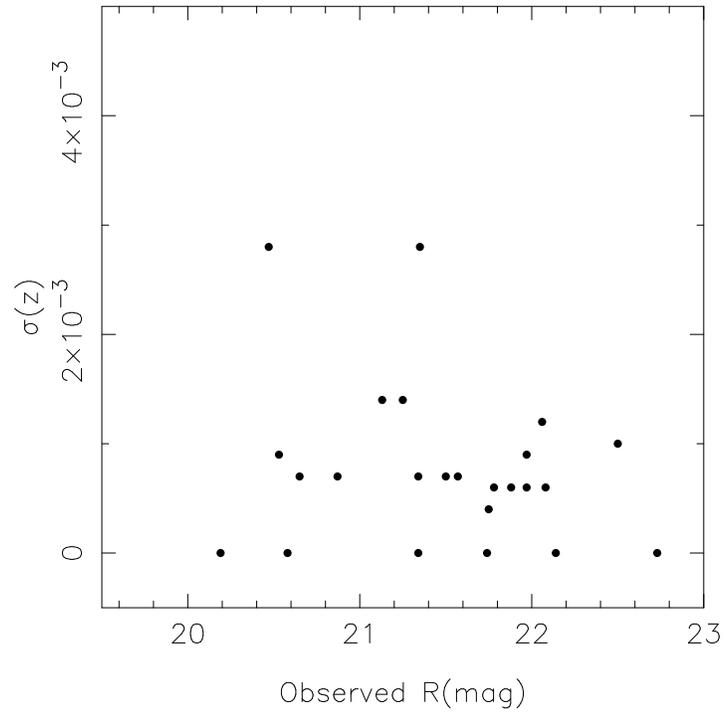}
\caption[figure1.ps]{The rms dispersion in redshift is shown
for galaxies with multiple spectra as a function of $R$ magnitude.
The implied scatter for a single redshift measurement is $\sim0.0008$.
\label{fig1}}
\end{figure}

\placefigure{fig1}

To demonstrate that the quality class 1 spectra (which comprise more
than 50\% of the extragalactic objects) actually do have the precision
claimed above, Figure~1 shows the values of $\sigma$ calculated from multiple
spectra of a single object where each individual measurement has a
quality 1 rating as a function of $R$ magnitude.  Only the galaxies
with multiple spectra
among the 60 brightest objects in the sample  were used.  There are four or
more spectra for three of these galaxies.

\section{Galaxy Spectroscopic Classifications \label{specclass} }

A simple spectral classification was used (Table~2).
``{$\cal E$}'' (emission) denotes an galaxy
where the emission lines dominate the spectrum, while the few extreme 
starburst galaxies that could be identified are denoted by ``{$\cal B$}''.
In the subsequent discussion, these two classes are usually
referred to collectively as
``{$\cal E$}'' (emission) galaxies.  ``{$\cal A$}'' (absorption)
denotes an object where no emission lines are detected, while
``{$\cal C$}'' (composite) is an intermediate case where both emission and
absorption (usually 3727 and H+K) are seen.  When both 3727 and
5007\AA\ emission lines are present, the object is denoted ``{$\cal E$}'', or,
if the absorption spectrum is also visible, ``{$\cal EC$}''.  
The three AGN are designated ``{$\cal Q$}''.  In addition,
stars are classified according to the presence or absence of TiO bands
(and/or CaH bands, for M subdwarfs)
as ``{$\cal M$}'' or ``{$\cal S$}''and, although there are no examples
in the present sample, as ``{$\cal W$}'' for white dwarfs.

\placetable{tab2}

The assignment of a spectral class to a galaxy will depend on its
redshift and on the wavelength range covered by the
observations.  For example, a galaxy at low redshift with strong
emission at 3727 and 5007\AA\ will be classified as ``{$\cal E$}'', 
but the same
object if seen at $z \sim 1.3$ when the 2800\AA\ region is observed in
the optical will show no apparent emission features and will be listed
as ``{$\cal A$}''.  This is particularly true for objects where the spectral
coverage of the data did not extend to $\lambda > 7500$\AA.
We emphasize that for $z < 0.9$, an ``{$\cal A$}'' galaxy
is an object that has a spectrum similar to that of local elliptical 
galaxies.   But at higher redshift, only the strongest emission lines
can be detected through the thicket of night sky lines, and a much
more diverse group of galaxies will be classified as  ``{$\cal A$}''
based on the presence of absorption lines in the 2500\AA\ region.

\begin{figure}
\epsscale{0.7}
\plotone{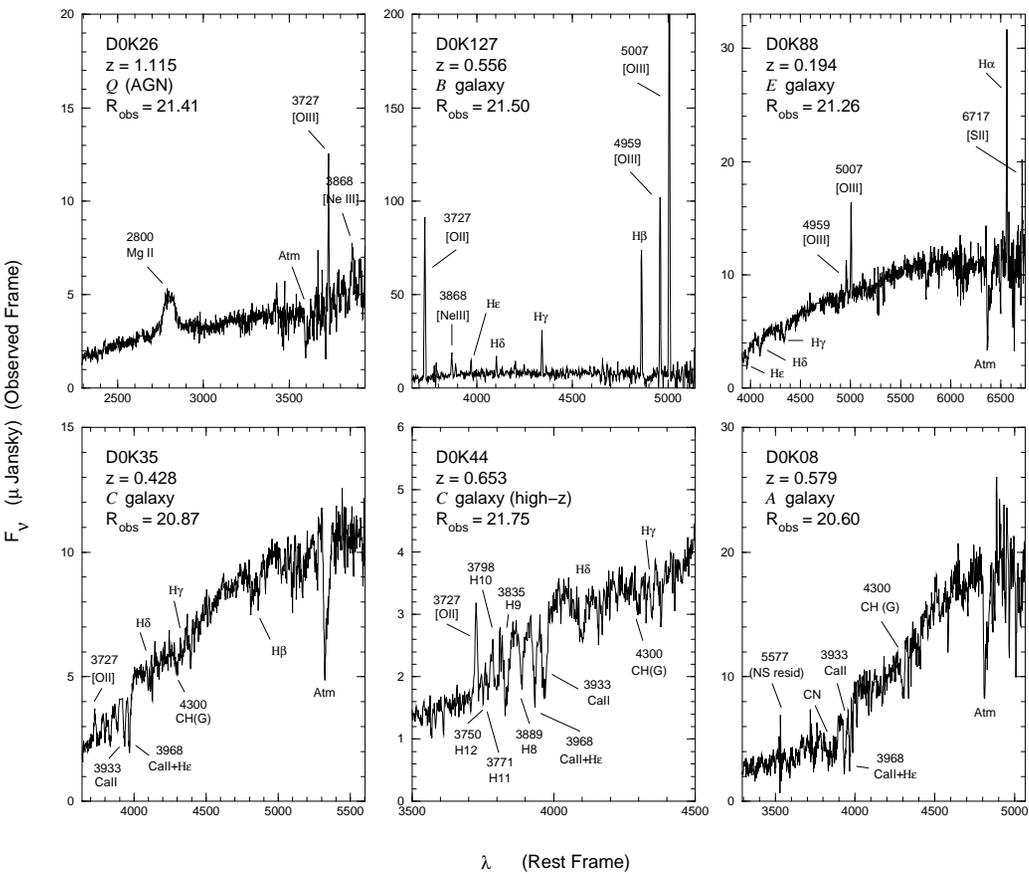}
\caption[figure2.eps]{Representative spectra of each of the five
spectral classes assigned to extragalactic objects are shown.  The
spectrum of the second brightest AGN and the second
brightest starburst galaxy is shown, while the fifth brightest
galaxy in $R$ is shown for each of the other three spectral classes.
The spectrum of a  high-$z$ ``${\cal C}$'' galaxy is also shown.
\label{fig2}}
\end{figure}

\placefigure{fig2}

Figure~2 gives illustrative examples of the five spectral classes used
for extragalactic objects.  The spectrum of the second brightest AGN 
and the second brightest starburst galaxy are shown,
while for the other three spectral classes, the spectrum of the
fifth brightest galaxy (at $R$) assigned to that class is shown. 
The observed $R$ magnitudes and $z$ are indicated for each object.
For the AGN and the spectral class 
``$\cal C$'' galaxy, the residuals
from sky subtraction around the very strong night sky
line at 5577\AA\ were removed by interpolation.  The ``$\cal C$'' galaxy
shown has a rather weak 3727\AA\ emission line compared to most
of the objects in this galaxy spectral class, but the distinction
between ``$\cal C$'' and ``$\cal A$'' spectral classes is still clear from the
relative strengths of absorption of H and K of CaII, the Balmer lines, 
and the molecular bands of CH and CN in this spectral region.  The
spectrum of a somewhat more typical higher-$z$ ``$\cal C$'' galaxy 
is also shown for comparison.

As will be discussed in Cohen \etal\ (1998\markcite{cohen98}),
there are many galaxies in this field with
$z \approx 0.58$.  At this redshift, the 3727\AA\ emission line of
[\ion{O}{2}], a key spectral type diagnostic, is observed at 5889\AA,
where it overlaps the strong NaD doublet in emission from the night
sky.  Unless the emission from such a galaxy at 3727\AA\ [\ion{O}{2}]
is very strong, it may be lost in the uncertainty of subtracting away
the much stronger night sky line.  Thus, for this redshift only, the
distinction between ``$\cal A$'' and ``$\cal C$'' spectral types has 
been made on the
basis of the presence of the 3880\AA\ CN band and the G band of CH at
4300\AA.  We assign such a galaxy a spectral class of ``$\cal A$'' if those
molecular bands are strong, while objects that show 4101\AA\ H$\delta$
absorption in their spectra are assigned the spectral class 
``$\cal C$''.  A
check of 19 objects originally classified as ``$\cal A$'' in this redshift
regime revealed three that had to be reclassified from ``$\cal A$'' to 
``$\cal C$''
on this basis as well as three others too faint to determine which
classification is most appropriate without the guidance of 3727\AA\
emission, which were left as ``$\cal A$''.

For the brighter part of this sample with high precision spectra, 
considerably finer distinction among galaxy spectral classes is possible.
A detailed discussion of the spectral features in the brighter galaxies
of this sample is deferred to a future paper 
(Cohen et al 1999\markcite{cohen99}).

\section{Redshifts for Objects in the Sample \label{redshifts} }

Table~3 lists the object numbers, positions on the sky, measured $K$
magnitudes (from \cite{pahre98apjs}), redshifts, spectral types, and
quality classes for 195 objects in the main sample, followed by the
13 additional objects that constitute the total sample.  Table~4 gives
the distribution of objects in the main sample over the spectral
and redshift quality classes.

\placetable{tab3}
\placetable{tab4}

Within the main sample, 84\% of the objects have measured redshifts.
The completeness is shown as a function of observed $R$ and of 
observed $K$ magnitude in
Figure~3.  As expected and supported by Figure~3, it is the $R$
magnitude that primarily determines whether or not a redshift can be
measured, rather than the $K$ magnitude. The median $R$ magnitude of the
objects never observed is 24.6~mag, while the median $R$ magnitude of the 21
objects that were observed, but for which no redshift was determined,
is 24.3 mag.  This latter group cannot contain any strong emission
line objects with $z < 1.1$ otherwise their spectra would have already
yielded redshifts.

All of the objects with redshifts in the main sample
have $R$, $I$ and $K$ magnitudes from the data of Pahre \etal\ (1998\markcite{pahre98apjs}), but several
are too faint to be detected in the available $U$, $B$, or $V$ images.

\begin{figure}
\epsscale{0.6}
\plotone{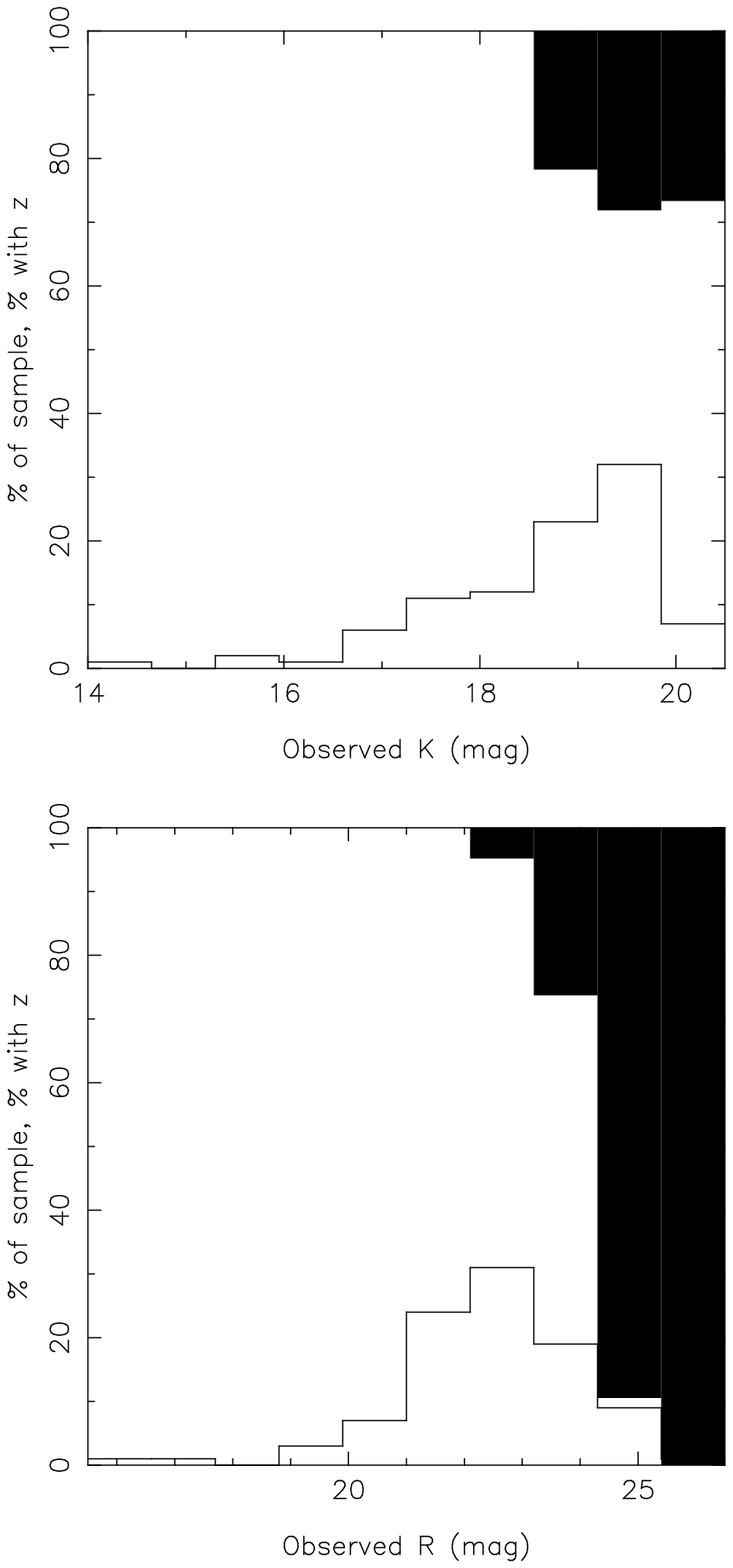}
\caption[figure3.ps]{A histogram of the sample and the completeness of the
redshifts are shown for the main sample.  The filled area denotes
the objects without redshifts. Figure~3a displays
this for the $K$ filter, while figure~3b show the results using the
$R$ filter.
\label{fig3}}
\end{figure}

\placefigure{fig3}

The median redshift of the extragalactic objects in the main sample
with measured $z$ is 0.58.  Even if it is assumed that all of the objects
without redshifts are galaxies with $z > 1$, then the median redshift can only
increase to $z_{med} = 0.65$.
Table~5 presents the median redshift for the extragalactic objects in
the main sample in half magnitude bins.  The objects without redshifts
have been ignored.

\placetable{tab5}

\section{Summary}

This paper provides a description of the techniques for analyzing and
characterizing
the spectra obtained by the Caltech Faint Galaxy Redshift Survey.
We have given here the basic observational results of our
spectroscopic investigation of a complete sample of objects 
with $K_s<20$~mag in a 2 by 7.3~arcmin field at J005325+1234.  
Redshifts were successfully obtained for 163 of the 195 objects in the sample
using the LRIS at the Keck Observatory.  An analysis of these data,
combined with the six-color photometric data of 
Pahre \etal (1998\markcite{pahre98apjs}),
for the extragalactic objects in this field will be given in
Cohen \etal\ (1998\markcite{cohen98}).

\acknowledgements The entire Keck/LRIS user community owes a huge debt
to Jerry Nelson, Gerry Smith, Bev Oke, and many other people who have
worked to make the Keck Telescope and LRIS a reality.  We are grateful
to the W. M. Keck Foundation, and particularly its late president,
Howard Keck, for the vision to fund the construction of the W. M. Keck
Observatory. JGC is grateful for partial support from STScI/NASA grant AR-06337.12-94A.
KR was supported in part by a Summer Undergraduate Research Fellowship
at Caltech.  RDB acknowledges support under NSF grant AST95-29170.
DWH and MAP were supported in part by Hubble Fellowship grants 
HF-01093.01-97A and HF-01099.01-97A from STScI (which is operated 
by AURA under NASA contract NAS5-26555).

\clearpage

%
%
\begin{deluxetable}{ll}
\tablenum{1}
\tablewidth{0pc}
\scriptsize
\tablecaption{Redshift Quality Classes}
\label{tab1}
\tablehead{
\colhead{Quality Class} & \colhead{Description of Class}}
\startdata
  1    &  multiple features, $\sigma(z) \le 0.002$/feature \nl
  2   &  multiple features, $\sigma(z) \le 0.004$/feature \nl
 3 &  multiple features, faint, id uncertain, \nl
  &  ~~~~$\sigma(z)$ small 75\% of time, and wildly off 25\% of time \nl
 4 & 1 emission line only, solid, assume 3727\AA \nl
 5 & 1 emission line only, reality uncertain, assume 3727\AA \nl
 6 & Multiple features, at least 1 broad emission line \nl
 7 & Only 1 broad emission line, assumed to be 2800\AA \nl
 8 & Single break, assumed to be 4000\AA\ break \nl
 9 & Single strong absorption feature, assumed to be 2800\AA \nl
   &  ~~~~because of shape of continuum \nl
 0 & No redshift \nl
\enddata
\end{deluxetable}

\clearpage

%
%
\begin{deluxetable}{cl}
\tablenum{2}
\tablewidth{0pc}
\scriptsize
\tablecaption{Definition of Spectral Types}
\label{tab2}
\tablehead{\colhead{Spectral Type} &  \colhead{Defining Features} }
\startdata 
\multicolumn{2}{c}{Galactic Stars:} \nl
$\cal M$ & TiO bands (M dwarfs); CaH bands (M subdwarfs) \nl
$\cal S$ & Absorption in Mg triplet, Balmer lines \nl
$\cal W$ & White dwarf, broad Balmer line absorption \nl
 &    \nl
\multicolumn{2}{c}{Extragalactic Objects:} \nl
$\cal Q$ & At least one broad emission line \nl
$\cal B$ & Emission in $H\delta$ and $H\epsilon$ as well as 3727, $H\beta$, 
     5007 \nl
$\cal E$ & Dominated by emission lines, 3727, 5007  \nl
$\cal C$ & composite, 3727 + Balmer line absorption \nl
$\cal A$ & No emission lines, only absorption features \nl
 &    \nl
\multicolumn{2}{c}{Unknown Objects:} \nl
$\cal F$ & Observed, but no redshift \nl
$\cal U$ & Never observed spectroscopically \nl
\enddata
\end{deluxetable}

%
%

%
%

\begin{deluxetable}{rrrcccc|lrrcccc}
\tablenum{3}
\tiny
\tablecolumns{14}
\tablewidth{0pc}
\tablecaption{Survey Objects}
\label{tab3}
\tablehead{
  \colhead{ID} &
  \colhead{RA\tablenotemark{1}} &
  \colhead{Dec} &
  \colhead{K} &
  \colhead{z} &
  \colhead{QC} &
  \colhead{SC} 
\vline &
  \colhead{ID} &
  \colhead{RA\tablenotemark{1}} &
  \colhead{Dec} &
  \colhead{K} &
  \colhead{z} &
  \colhead{QC} &
  \colhead{SC}  \\
  \colhead{[D0K]} &
  \colhead{[$-0^{\rm h}$]} &
  \colhead{[$-12$\arcdeg]} &
  \colhead{[mag]} &
  \colhead{} &
  \colhead{} &
  \colhead{}  \vline &
  \colhead{[D0K]} &
  \colhead{[$-0^{\rm h}$]} &
  \colhead{[$-12$\arcdeg]} &
  \colhead{[mag]} &
  \colhead{} &
  \colhead{} &
  \colhead{} 
}
\startdata
   1 & 5327.93 & 3018.40 &   12.73 &   0.000 &  1 &  $\cal M $   &    2 & 5327.69 & 3214.60 &   14.43 &   0.000 &  1 &  $\cal S $  \nl
   3 & 5322.88 & 3546.60 &   14.91 &   0.000 &  1 &  $\cal S $   &    4 & 5322.94 & 3352.90 &   15.38 &   0.000 &  1 &  $\cal M $  \nl
   5 & 5328.48 & 3717.70 &   15.58 &   0.000 &  1 &  $\cal M $   &    6 & 5328.72 & 3029.50 &   15.68 &   0.428 &  1 &  $\cal A $  \nl
   7 & 5328.48 & 3731.40 &   15.87 &   0.000 &  1 &  $\cal S $   &    8 & 5327.87 & 3332.50 &   16.27 &   0.579 &  1 &  $\cal A $  \nl
   9 & 5327.94 & 3613.70 &   16.30 &   0.441 &  1 &  $\cal C $   &   10 & 5325.61 & 3718.90 &   16.49 &   0.000 &  1 &  $\cal M $  \nl
  11 & 5323.82 & 3729.90 &   16.67 &   0.346 &  1 &  $\cal A $   &   12 & 5324.73 & 3342.40 &   16.75 &   0.681 &  6 &  $\cal Q $  \nl
  13 & 5324.12 & 3027.40 &   16.83 &   0.588 &  1 &  $\cal C $   &   14 & 5325.89 & 3536.90 &   16.88 &   0.428 &  1 &  $\cal A $  \nl
  15 & 5324.55 & 3732.00 &   16.89 &   0.000 &  1 &  $\cal M $   &   16 & 5328.16 & 3044.40 &   16.95 &   0.431 &  1 &  $\cal A $  \nl
  17 & 5322.22 & 3209.90 &   16.98 &   0.392 &  1 &  $\cal A $   &   18 & 5322.56 & 3252.30 &   17.02 &   0.173 &  1 &  $\cal C $  \nl
  19 & 5325.95 & 3151.20 &   17.02 &   0.581 &  1 &  $\cal A $   &   20 & 5324.71 & 3437.80 &   17.10 &   0.000 &  1 &  $\cal M $  \nl
  21 & 5327.85 & 3652.30 &   17.13 &   0.000 &  1 &  $\cal M $   &   22 & 5325.87 & 3144.90 &   17.14 &   0.581 &  1 &  $\cal A $  \nl
  23 & 5323.46 & 3415.40 &   17.21 &   0.000 &  1 &  $\cal M $   &   24 & 5325.94 & 3423.30 &   17.30 &   0.679 &  1 &  $\cal C $  \nl
  25 & 5324.24 & 3639..40 &   17.39 &   0.309 &  1 &  $\cal C $   &   26 & 5322.37 & 3059.00 &   17.44 &   1.115 &  6 &  $\cal Q $  \nl
  27 & 5326.13 & 3147.80 &   17.44 &   0.577 &  1 &  $\cal A $   &   28 & 5324.43 & 3409.10 &   17.44 &   0.582 &  1 &  $\cal A $  \nl
  29 & 5326.30 & 3239.20 &   17.45 &   0.582 &  1 &  $\cal AC $   &   30 & 5328.38 & 3501.80 &   17.52 &   0.621 &  1 &  $\cal C $  \nl
  31 & 5329.12 & 3222.70 &   17.53 &   0.430 &  1 &  $\cal A $   &   32 & 5329.32 & 3330.10 &   17.54 &   0.577 &  1 &  $\cal A $  \nl
  33 & 5325.91 & 3558.40 &   17.54 &   0.429 &  1 &  $\cal A $   &   34 & 5324.73 & 3411.80 &   17.55 &   0.679 &  1 &  $\cal AC $  \nl
  35 & 5329.07 & 3457.20 &   17.56 &   0.428 &  1 &  $\cal C $   &   36 & 5323.73 & 3711.00 &   17.58 &   0.605 &  1 &  $\cal AC$  \nl
  37 & 5326.88 & 3634.40 &   17.69 &   0.772 &  1 &  $\cal A $   &   38 & 5327.08 & 3610.40 &   17.75 &   0.763 &  1 &  $\cal A $  \nl
  39 & 5323.21 & 3328.60 &   17.75 &   0.583 &  1 &  $\cal C $   &   40 & 5324.94 & 3202.30 &   17.80 &   0.582 &  1 &  $\cal A $  \nl
  41 & 5322.00 & 3303.90 &   17.82 &   0.654 &  1 &  $\cal AC$   &   42 & 5327.65 & 3522.40 &   17.82 &   1.136 &  3 &  $\cal A $  \nl
  43 & 5321.79 & 3018.00 &   17.86 &   0.000 &  1 &  $\cal M $   &   44 & 5321.92 & 3320.30 &   17.88 &   0.653 &  1 &  $\cal C $  \nl
  45 & 5325.90 & 3158.70 &   17.88 &   0.581 &  1 &  $\cal A $   &   46 & 5325.50 & 3047.30 &   17.91 &   0.763 &  1 &  $\cal C $  \nl
  47 & 5324.59 & 3347.00 &   17.99 &   0.582 &  1 &  $\cal C $   &   48 & 5327.76 & 3545.30 &   18.00 &   0.578 &  2 &  $\cal C $  \nl
  49 & 5323.74 & 3441.70 &   18.06 &   0.584 &  1 &  $\cal C $   &   50 & 5323.10 & 3234.50 &   18.08 &   0.000 &  1 &  $\cal M $  \nl
  51 & 5321.54 & 3135.40 &   18.09 &   0.677 &  1 &  $\cal C $   &   52 & 5327.78 & 3457.30 &   18.10 &   0.680 &  1 &  $\cal C $  \nl
  53 & 5329.39 & 3307.20 &   18.10 &   0.000 &  1 &  $\cal S $   &   54 & 5328.49 & 3724.10 &   18.11 &   0.209 &  1 &  $\cal EC$   \nl
  55 & 5324.02 & 3336.60 &   18.13 &   0.432 &  1 &  $\cal C $   &   56 & 5325.90 & 3229.60 &   18.27 &   0.689 &  3 &  $\cal A $  \nl
  57 & 5327..08 & 3132.00 &   18.28 &   0.369 &  1 &  $\cal EB$   &   58 & 5328.86 & 3251.90 &   18.28 &   0.509 &  1 &  $\cal C $  \nl
  59 & 5322.50 & 3549.70 &   18.29 &   0.771 &  1 &  $\cal C $   &   60 & 5329.13 & 3719.00 &   18.30 &   0.781 &  3 &  $\cal A $  \nl
  61 & 5326.73 & 3650.20 &   18.32 &   0.000 &  1 &  $\cal S $   &   62 & 5327.54 & 3417.70 &   18.34 &   0.584 &  2 &  $\cal A $  \nl
  63 & 5321.73 & 3141.00 &   18.35 &   0.535 &  3 &  $\cal C $   &   64 & 5324.50 & 3701.90 &   18.42 &   1.048 &  3 &  $\cal A $  \nl
  65 & 5326.91 & 3051.30 &   18.44 &   1.232 &  3 &  $\cal A $   &   66 & 5325.60 & 3243.90 &   18.46 &   0.771 &  1 &  $\cal EC$  \nl
  67 & 5327.95 & 3255.90 &   18.46 &   0.584 &  1 &  $\cal A $   &   68 & 5321.40 & 3153.70 &   18.47 &   1.392 &  2 &  $\cal A $  \nl
  69 & 5321.32 & 3514.40 &   18.51 &   1.336 &  3 &  $\cal A $   &   70 & 5328.97 & 3156.00 &   18.54 &   0.581 &  1 &  $\cal A $  \nl
  71 & 5323.70 & 3634.50 &   18.54 &   0.209 &  1 &  $\cal E $   &   72 & 5321.96 & 3341.10 &   18.56 &    ...  &  0 &  $\cal F $  \nl
  73 & 5326.51 & 3419.00 &   18.57 &   0.493 &  1 &  $\cal C $   &   74 & 5321.62 & 3117.90 &   18.60 &   0.676 &  2 &  $\cal C $  \nl
  75 & 5322.20 & 3658.30 &   18.60 &   0.763 &  1 &  $\cal C $   &   76 & 5322.76 & 3155.00 &   18.63 &   1.153 &  6 &  $\cal Q $  \nl
  77 & 5325.63 & 3510.80 &   18.64 &    ...  &  0 &  $\cal F $   &   78 & 5329.70 & 3104.30 &   18.66 &   0.633 &  1 &  $\cal B $  \nl
  79 & 5321.35 & 3346.80 &   18.68 &   0.533 &  1 &  $\cal A $   &   80 & 5325.68 & 3444.50 &   18.74 &   0.585 &  2 &  $\cal C $  \nl
  81 & 5326.81 & 3030.10 &   18.75 &   0.414 &  1 &  $\cal C $   &   82 & 5323.64 & 3343.70 &   18.76 &   0.000 &  1 &  $\cal M $  \nl
  83 & 5321.70 & 3426.30 &   18.80 &   0.429 &  1 &  $\cal C $   &   84 & 5326.45 & 3359.20 &   18.81 &   0.678 &  1 &  $\cal C $  \nl
  85 & 5326.15 & 3350.40 &   18.81 &   0.626 &  1 &  $\cal C $   &   86 & 5324.44 & 3515.60 &   18.82 &   0.549 &  1 &  $\cal E $  \nl
  87 & 5327.41 & 3308.50 &   18.82 &   0.440 &  1 &  $\cal E $   &   88 & 5327.26 & 3627.90 &   18.85 &   0.194 &  1 &  $\cal E $  \nl
  89 & 5325.48 & 3108.40 &   18.86 &   0.581 &  1 &  $\cal A $   &   90 & 5328.99 & 3442.40 &   18.87 &   0.511 &  1 &  $\cal C $  \nl

\tablebreak
  91 & 5324.24 & 3709.00 &   18.88 &    ...  &  0 &  $\cal F $   &   92 & 5324.11 & 3645.90 &   18.89 &   0.000 &  1 &  $\cal M $  \nl
  93 & 5322.02 & 3216.60 &   18.90 &   1.043 &  2 &  $\cal A $   &   94 & 5322.52 & 3732.00 &   18.91 &    ...  &  0 &  $\cal F $  \nl
  95 & 5326.78 & 3626.20 &   18.91 &   0.000 &  1 &  $\cal M $   &   96 & 5324.18 & 3020.20 &   18.94 &   0.677 &  1 &  $\cal C $  \nl
  97 & 5328.80 & 3434.70 &   18.94 &   0.531 &  8 &  $\cal A $   &   98 & 5322.04 & 3045.50 &   18.97 &    ...  &  0 &  $\cal F $  \nl
  99 & 5324.95 & 3035.40 &   18.98 &   0.000 &  1 &  $\cal M $   &  100 & 5326.95 & 3552.90 &   19.00 &   0.680 &  2 &  $\cal C $  \nl
 101 & 5325.93 & 3214.10 &   19.01 &   0.578 &  1 &  $\cal A $   &  102 & 5323.44 & 3336.00 &   19.01 &   0.448 &  3 &  $\cal A $  \nl
 103 & 5321.72 & 3038.80 &   19.01 &   0.607 &  1 &  $\cal EC$   &  104 & 5323.46 & 3558.00 &   19.05 &   0.182 &  1 &  $\cal B $  \nl
 105 & 5328.83 & 3334.80 &   19.05 &   0.681 &  1 &  $\cal E $   &  106 & 5322.29 & 3646.70 &   19.06 &    ...  &  0 &  $\cal U $  \nl
 107 & 5324.77 & 3148.00 &   19.10 &    ...  &  0 &  $\cal F $   &  108 & 5328.14 & 3527.30 &   19.13 &   1.013 &  1 &  $\cal E $  \nl
 109 & 5321.56 & 3607.60 &   19.15 &   0.000 &  1 &  $\cal M $   &  110 & 5327.24 & 3048.70 &   19.15 &   0.391 &  1 &  $\cal E $  \nl
 111 & 5327.27 & 3136.90 &   19.17 &   0.582 &  1 &  $\cal C $   &  112 & 5324.19 & 3230.00 &   19.19 &   1.298 &  2 &  $\cal EC$  \nl
 113 & 5326.12 & 3428.50 &   19.19 &   1.264 &  3 &  $\cal A $   &  114 & 5329.23 & 3447.00 &   19.21 &   0.583 &  1 &  $\cal C $  \nl
 115 & 5327.94 & 3559.20 &   19.21 &    ...  &  0 &  $\cal F $   &  116 & 5326.62 & 3432.50 &   19.24 &   0.584 &  1 &  $\cal A $  \nl
 117 & 5328.22 & 3721.70 &   19.26 &    ...  &  0 &  $\cal F $   &  118 & 5327.88 & 3429.80 &   19.27 &    ...  &  0 &  $\cal U $  \nl
 119 & 5326.11 & 3347.50 &   19.27 &    ...  &  0 &  $\cal F $   &  120 & 5325.79 & 3125.60 &   19.27 &    ...  &  0 &  $\cal F $  \nl
 121 & 5325.47 & 3535.00 &   19.28 &   0.948 &  1 &  $\cal E $   &  122 & 5323.85 & 3413.40 &   19.30 &   0.570 &  2 &  $\cal C $  \nl
 123 & 5322.47 & 3116.80 &   19.30 &   1.440 &  4 &  $\cal E $   &  124 & 5322.09 & 3037.40 &   19.30 &    ...  &  0 &  $\cal U $  \nl
 125 & 5323.10 & 3253.10 &   19.31 &   1.009 &  4 &  $\cal E $   &  126 & 5324.27 & 3017.20 &   19.31 &   1.301 &  5 &  $\cal E $  \nl
 127 & 5323.24 & 3350.60 &   19.33 &   0.556 &  1 &  $\cal B $   &  128 & 5322.88 & 3551.10 &   19.34 &   1.307 &  3 &  $\cal A $  \nl
 129 & 5329.21 & 3537.40 &   19.35 &    ...  &  0 &  $\cal U $   &  130 & 5320.96 & 3318.20 &   19.35 &    ...  &  0 &  $\cal F $  \nl
 131 & 5323.83 & 3312.00 &   19.36 &   0.580 &  2 &  $\cal C $   &  132 & 5323.76 & 3215.40 &   19.36 &   0.391 &  1 &  $\cal E $  \nl
 133 & 5323.64 & 3234.80 &   19.38 &   0.582 &  1 &  $\cal C $   &  134 & 5321.16 & 3120.10 &   19.39 &   0.000 &  1 &  $\cal M $  \nl
 135 & 5324.48 & 3224.70 &   19.40 &   1.223 &  3 &  $\cal A $   &  136 & 5324.83 & 3430.40 &   19.40 &    ...  &  0 &  $\cal F $  \nl
 137 & 5323.66 & 3335.90 &   19.42 &   0.872 &  1 &  $\cal EC$   &  138 & 5322.55 & 3502.00 &   19.43 &   0.428 &  1 &  $\cal C $  \nl
 139 & 5327.35 & 3245.10 &   19.43 &   0.573 &  3 &  $\cal A $   &  140 & 5328.40 & 3647.70 &   19.44 &   0.772 &  1 &  $\cal C $  \nl
 141 & 5321.35 & 3124.30 &   19.45 &    ...  &  0 &  $\cal F $   &  142 & 5322.40 & 3627.80 &   19.46 &   0.761 &  1 &  $\cal A $  \nl
 143 & 5321.47 & 3319.80 &   19.47 &   0.750 &  3 &  $\cal A $   &  144 & 5323.55 & 3430.00 &   19.51 &   0.370 &  1 &  $\cal E $  \nl
 145 & 5321.31 & 3158.60 &   19.51 &   0.581 &  1 &  $\cal C $   &  146 & 5329.14 & 3551.90 &   19.51 &    ...  &  0 &  $\cal F $  \nl
 147 & 5326.38 & 3116.20 &   19.52 &   0.415 &  1 &  $\cal C $   &  148 & 5328.79 & 3508.10 &   19.52 &    ...  &  0 &  $\cal U $  \nl
 149 & 5323.18 & 3711.10 &   19.53 &   0.269 &  1 &  $\cal EB$   &  150 & 5321.73 & 3724.30 &   19.53 &   0.862 &  3 &  $\cal C $  \nl
 151 & 5322.98 & 3141.10 &   19.53 &   0.492 &  2 &  $\cal C $   &  152 & 5328.83 & 3103.40 &   19.55 &   1.075 &  9 &  $\cal A $  \nl
 153 & 5326.38 & 3412.10 &   19.61 &   0.495 &  1 &  $\cal C $   &  154 & 5329.14 & 3542.20 &   19.62 &    ...  &  0 &  $\cal U $  \nl
 155 & 5321.35 & 3730.20 &   19.63 &    ...  &  0 &  $\cal F $   &  156 & 5324.38 & 3241.30 &   19.63 &    ...  &  0 &  $\cal F $  \nl
 157 & 5323.71 & 3609.30 &   19.63 &   0.535 &  1 &  $\cal C $   &  158 & 5328.04 & 3140.90 &   19.65 &   1.306 &  1 &  $\cal E $  \nl
 159 & 5321.43 & 3256.00 &   19.66 &   0.394 &  1 &  $\cal E $   &  160 & 5327.69 & 3607.30 &   19.68 &   0.000 &  1 &  $\cal M $  \nl
 161 & 5324.43 & 3626.70 &   19.70 &   0.432 &  1 &  $\cal C $   &  162 & 5322.76 & 3644.80 &   19.71 &   0.643 &  1 &  $\cal C $  \nl
 163 & 5324.72 & 3356.00 &   19.72 &   0.580 &  3 &  $\cal A $   &  164 & 5328.66 & 3139.30 &   19.72 &   0.587 &  2 &  $\cal C $  \nl
 165 & 5328.48 & 3056.80 &   19.73 &   0.581 &  8 &  $\cal A $   &  166 & 5326.60 & 3547.70 &   19.74 &   0.611 &  1 &  $\cal EC$  \nl
 167 & 5329.19 & 3351.70 &   19.74 &   0.442 &  1 &  $\cal C $   &  168 & 5321.60 & 3204.00 &   19.74 &    ...  &  0 &  $\cal U $  \nl
 169 & 5325.51 & 3530.70 &   19.76 &    ....  &  0 &  $\cal F $   &  170 & 5322.02 & 3547.60 &   19.76 &   0.533 &  1 &  $\cal E $  \nl
 171 & 5326.33 & 3020.90 &   19.77 &   0.345 &  1 &  $\cal E $   &  172 & 5322.74 & 3209.00 &   19.77 &   0.974 &  9 &  $\cal A $  \nl
 173 & 5321.11 & 3243.50 &   19.81 &    ...  &  0 &  $\cal U $   &  174 & 5325.24 & 3640.70 &   19.83 &  0.968  &  5 &  $\cal E $  \nl
 175 & 5325.66 & 3434.70 &   19.84 &   0.939 &  4 &  $\cal E $   &  176 & 5324.05 & 3536.10 &   19.85 &    ...  &  0 &  $\cal F $  \nl
 177 & 5325.52 & 3559.00 &   19.86 &    ...  &  0 &  $\cal U $   &  178 & 5326.15 & 3359.60 &   19.88 &   0.745 &  3 &  $\cal A $  \nl
 179 & 5328.56 & 3618.20 &   19.89 &   0.763 &  2 &  $\cal C $   &  180 & 5322.23 & 3650.60 &   19.90 &    ...  &  0 &  $\cal F $  \nl

\tablebreak
 181 & 5324.83 & 3254.70 &   19.91 &   1.040 &  9 &  $\cal A $   &  182 & 5323.55 & 3357.60 &   19.92 &   1.137 &  3 &  $\cal A $  \nl
 183 & 5323.03 & 3437.60 &   19.92 &   0.922 &  1 &  $\cal E $   &  184 & 5322.63 & 3509.80 &   19.92 &   0.622 &  1 &  $\cal C $  \nl
 185 & 5327.51 & 3611.40 &   19.92 &   0.975 &  1 &  $\cal E $   &  186 & 5321.63 & 3020.50 &   19.95 &    ...  &  0 &  $\cal U $  \nl
 187 & 5324.24 & 3630.60 &   19.96 &   0.344 &  1 &  $\cal C $   &  188 & 5324.47 & 3617.90 &   19.98 &   1.218 &  2 &  $\cal A $  \nl
 189 & 5321.69 & 3303.70 &   19.98 &   0.490 &  4 &  $\cal A $   &  450 & 5329.09 & 3030.30 &   18.12 &   0.000 &  1 &  $\cal M $  \nl
 956 & 5325.20 & 3615.60 &   19.48 &   0.000 &  1 &  $\cal M $   &  494 & 5326.02 & 3534.90 &   18.91 &    ...  &  0 &  $\cal F $  \nl
 504 & 5322.56 & 3252.30 &   19.06 &    ...  &  0 &  $\cal F $   &  506 & 5322.88 & 3546.60 &   19.07 &    ...  &  0 &  $\cal U $  \nl
 507 & 5324.73 & 3408.00 &   19.12 &  1.002  &  4 &  $\cal E $   \nl

\cutinhead{Supplemental objects}
 451 & 5322.12 & 3354.90 &   20.65 &   0.430 &  8 &  $\cal A $   &  951 & 5329.56 & 3644.50 &  -20.00 &   0.440 &  1 &  $\cal C $  \nl
 953 & 5330.64 & 3632.40 &  -20.00 &   0.280 &  1 &  $\cal E $   &  958 & 5327.06 & 3546.90 &   20.02 &   0.442 &  1 &  $\cal C $  \nl
 961 & 5318.41 & 3516.40 &  -20.00 &   0.000 &  1 &  $\cal S $   &  968 & 5324.21 & 3325.70 &   20.58 &   0.000 &  1 &  $\cal S $  \nl
 971 & 5331.26 & 3212.00 &  -20.00 &   0.352 &  1 &  $\cal E $   &  973 & 5330.40 & 3041.00 &  -20.00 &   0.340 &  1 &  $\cal E $  \nl
 980 & 5326.00 & 3157.01 &   20.20 &   0.437 &  1 &  $\cal E $   &  982 & 5329.11 & 3020.81 &   20.45 &   1.420 &  9 &  $\cal A $  \nl
 983 & 5325.62 & 3504.53 &  -20.00 &   0.460 &  8 &  $\cal A $   &  984 & 5321.50 & 3123.00 &   20.60 &   1.119 &  1 &  $\cal E $  \nl
 985 & 5321.16 & 3004.50 &  -20.00 &   0.773 &  4 &  $\cal E $  \nl

\enddata
\tablenotetext{1}{ID, RA, Dec, and R magnitude from Pahre \etal\ 1998.}

\tablecomments{QC is the quality code described in \S\ref{qualityclasses}.  
SC is the spectral classification described in \S\ref{specclass}.}

\end{deluxetable}

\clearpage

%
%
\begin{deluxetable}{crcr}
\tablenum{4}
\tablewidth{0pc}
\scriptsize
\tablecaption{Distribution of Sample Among Galaxy Spectral Types and Quality 
Classes}
\label{tab4}
\tablehead{\colhead{Spectral Class } & \colhead{No. of Objects} &
\colhead{Quality Class} & \colhead{No. of Objects} }
\startdata
$\cal M$ & 19 & 1 & 117\tablenotemark{1} \nl
$\cal S$ &  5 & 2 & 14 \nl
$\cal Q$ &  3 & 3 & 17 \nl
$\cal B$ &  3 & 4 & 5 \nl
$\cal E$ & 30 & 5 & 2 \nl
$\cal C$ & 50 & 6 & 3 \nl
$\cal A$ & 53 & 7 & 0 \nl
$\cal F$ & 21 & 8 & 2 \nl
$\cal U$ & 11 & 9 & 3 \nl
         &    & 0 & 32 \nl
\enddata
\tablenotetext{1}{Includes 24 Galactic stars.}
\end{deluxetable}

%
%
\begin{deluxetable}{crrrr}
\tablenum{5}
\tablewidth{0pc}
\scriptsize
\tablecaption{Median Redshift with Magnitude}
\label{tab5}
\tablehead{\colhead{magnitude} & \multicolumn{2}{c}{$R$} &
\multicolumn{2}{c}{$K$} \nl \colhead{bin} & \colhead{$N$} &
\colhead{$\left< z \right>$} & \colhead{$N$} & \colhead{$\left< z
\right>$}}
\startdata
15.5-16.0 &  0 &      &  1 & 0.43 \nl
16.0-16.5 &  0 &      &  2 & 0.44 \nl
16.5-17.0 &  0 &      &  6 & 0.43 \nl
17.0-17.5 &  0 &      &  9 & 0.58 \nl
17.5-18.0 &  0 &      & 18 & 0.58 \nl
18.0-18.5 &  0 &      & 17 & 0.58 \nl
18.5-19.0 &  1 & 0.68 & 23 & 0.58 \nl
19.0-19.5 &  2 & 0.17 & 30 & 0.61 \nl
19.5-20.0 &  1 & 0.31 & 33 & 0.58 \nl
20.0-20.5 &  7 & 0.43 &    &      \nl
20.5-21.0 &  4 & 0.43 &    &      \nl
21.0-21.5 & 23 & 0.44 &    &      \nl
21.5-22.0 & 23 & 0.58 &    &      \nl
22.0-22.5 & 21 & 0.58 &    &      \nl
22.5-23.0 & 28 & 0.64 &    &      \nl
23.0-23.5 & 20 & 0.86 &    &      \nl
23.5-24.0 &  5 & 0.76 &    &      \nl
24.0-24.5 &  4 & 0.75 &    &      \nl
\enddata
\end{deluxetable}

\clearpage



\clearpage

%
%

\end{document}